\documentclass[a4paper,10pt,twoside]{cpc-hepnp}

\usepackage{multicol}
\usepackage{graphicx}
\usepackage{booktabs}
\usepackage{amssymb,bm,mathrsfs,bbm,amscd}
\usepackage[tbtags]{amsmath}
\usepackage{lastpage}

\begin{document}

\fancyhead[co]{\footnotesize R. A. Arndt et al: Baryon Resonance Analysis from SAID}

\footnotetext[0]{Received 14 June 2009}

\title{Baryon Resonance Analysis from SAID\thanks{Supported by the U.~S.~Department 
of Energy under Grants  (DE--FG02--99ER41110 and DE--AC05--84ER40150) }}

\author{%
      R. A. Arndt%
\quad W. J. Briscoe%
\quad M. W. Paris%
\quad I. I. Strakovsky$^1)$\email{igor@gwu.edu}%
\quad R. L. Workman
}
\maketitle

\address{%
Center for Nuclear Studies, Department of Physics, \\
The George Washington University, Washington, D.C. 20052, U.S.A.
\\
}

\begin{abstract}
We discuss the analysis of data from $\pi$N elastic scattering and
single pion photo- and electroproduction.  The main focus is a
study of low-lying non-strange baryon resonances. Here we concentrate
on some difficulties associated with resonance identification, 
in particular the Roper and higher P$_{11}$ states.
\end{abstract}



\begin{multicols}{2}

\section{Introduction}

Many of the SAID fits to scattering data have been motivated by ongoing
studies of the $N^\ast$ properties\cite{PDG}. Most of these (for 
instance, EBAC\cite{EBAC}, Giessen\cite{Giessen}, DMT\cite{DMT},
J\"ulich\cite{Juelich}) have used, as input, amplitudes extracted 
from elastic $\pi$N scattering data\cite{piN,piNo}. Our pion 
photoproduction multipoles are also determined using a $K$-matrix 
formalism based upon $\pi$N partial-wave 
amplitudes\cite{piPR,piPRn,piPRp}. Further, 
the pion-electroproduction analysis is anchored to our $Q^2 = 0$ 
photoproduction results, with additional factors intended to account 
for the $Q^2$ variation\cite{piEPR}.

One of the most convincing ways to study the spectroscopy of 
non-strange baryons is through $\pi$N partial-wave analysis (PWA).  
The main sources of the Review of Particle Physics (RPP) $N^\ast$ 
Listings\cite{PDG} are the PWA of the KH, CMB, and GW/VPI groups.  
The analysis of $\pi$N scattering data remains crucial in this 
effort. Double-polarization quantities (R and A) measured long
after the KH and CMB analyses were completed have found discrepancies
in these earlier fits, which weakens claims for the existence and
properties of some of the weaker (mainly isospin 3/2) resonances.  

In the GW DAC $\pi$N PWA, we determine $\pi$N amplitudes by the fitting 
$\pi$N elastic data (up to W = 2.50~GeV) and $\pi^-p\to\eta n$ data (up 
to W = 1.63~GeV).  Resonances are then found through a search for poles 
in the complex energy plane.  We consider mainly poles which are not far 
away from the physical axis.  It is important to emphasize that these 
resonances are not put in by hand, contrary to the Breit-Wigner (BW) 
parametrization. The poles arise, in a sense, dynamically as a result of 
the enforced (quasi-) two-body unitarity cuts and the fit to the observable 
on the real energy axis. We have, however, also given the results of a BW 
parametrization, mapping $\chi^2$[W$_R$, $\Gamma$] while searching all 
other partial-wave parameters by fitting data over a relatively narrow 
energy range, say 100--200~MeV. Some subjectivity in the BW study is 
involved, such as: 
(i) energy binning, 
(ii) the strength of constraints (such as dispersion relations), and 
(iii) the choice of partial waves to be searched.  
We should stress that the standard PWA reveals resonances with 
widths of order 100 MeV, but not too wide ($\Gamma >$ 500~MeV) or 
possessing too small a branching ratio (BR $<$ 4\%), tending (by 
construction) to miss narrow resonances with $\Gamma <$ 30~MeV.  
The partial waves of solution KA84\cite{KH} and the single-energy 
solutions (SES) associated with our SP06 results agree reasonably 
well over the full energy range of the SP06 (Figs.~4--7 
from\cite{piN}).  However, this does not lead to agreement on the 
resonance content. For instance, our study \cite{piN} does not 
support several N$^\ast$ and $\Delta^\ast$ reported by PDG\cite{PDG}.  
It is important here to remember that during last 25 years, the $\pi$N 
database has increased by a factor of 3--4, and these data were not
available to the KH and CMB groups.

\section{$\pi$N Features}

\subsection{Minimization and Normalization factor}

As in previous analyses, we have used the systematic uncertainty as 
an overall normalization factor for angular distributions.  
Renormalization freedom 
significantly improves our best-fit results, as shown in 
Table~\ref{tab1} (we use the same methodology in all of our PWAs).  This 
renormalization procedure was also applied to the other non-SAID 
solutions. Here, however, only the normalization constants were 
searched to minimize $\chi^2$ (no adjustment of the partial waves 
was possible).  Clearly, this procedure can significantly improve 
the overall $\chi^2$ attributed to a fit (we cannot ignore this
experimental input), and has been applied in calculating the 
$\chi^2$ values of Table~\ref{tab1}.
\end{multicols}
\begin{center}
\tabcaption{ \label{tab1} Comparison of $\chi^2$/data values 
         (Norm/Unnorm) for normalized
         (Norm) and unnormalized (Unnorm) data used in the 
         SP06\protect\cite{piN} and FA02\protect\cite{piNo} solutions,  
         Karlsruhe KA84\protect\cite{KH}, EBAC\protect\cite{EBAC},
         Giessen\protect\cite{Giessen}, and DMT\protect\cite{DMT}.  
	 Values for SP06 (FA02) correspond to a 2.46 (2.26)~GeV energy 
         limit for W in CM.  KA84 is evaluated up to 2.9~GeV, EBAC up to 
         1.91~GeV, Giessen up to 2~GeV, and DMT up to 2.2~GeV.} 
\vspace{3mm}
\footnotesize
\begin{tabular*}{170mm}{c@{\extracolsep{\fill}}cccccc}
\toprule Reaction &      SP06      &       FA02      &      KA84      &      EBAC      &    Giessen     &    DMT    \\
\toprule          &  $\chi^2$/Data &   $\chi^2$/Data &  $\chi^2$/Data &  $\chi^2$/Data &  $\chi^2$/Data & $\chi^2$/Data\\
\hline
$\pi^+p\to\pi^+p$ &    2.0/6.1     &     2.1/8.8     &     5.0/24.9   &      13.1/23.7 &     10.5/17.7  &   15.4/37.4\\
$\pi^-p\to\pi^-p$ &    1.9/6.2     &     2.0/6.6     &     9.1/51.9   &      4.9/16.0  &     12.1/34.1  &    9.0/23.0\\
$\pi^-p\to\pi^0n$ &    2.0/4.0     &     1.9/5.9     &     4.4/8.8    &      3.5/6.3   &     6.3/15.2   &    6.5/16.7\\
$\pi^-p\to\eta n$ &    2.5/9.6     &     2.5/10.5    &                &                &                &    \\
\bottomrule
\end{tabular*}
\end{center}

\begin{multicols}{2}
\subsection{Roper}

Discovered more than 40 years ago\cite{Roper}, this resonance state has 
remained controversial for many years.  The prominent N(1440)P$_{11}$ 
resonance is clearly evident in both KH and GW/VPI analyses 
(Figs.~4--7 from Ref.\cite{piN}), but occurs very near the $\pi\Delta$, 
$\eta$N, and $\rho$N thresholds (Fig.~8 from Ref.\cite{piNo}), making a 
BW fit questionable.  The N(1440) is unique in that its behavior on
the real energy axis is influenced by poles on different Riemann sheets 
(with respect to the $\pi\Delta$-cut) as was first reported 
by Arndt et al.\cite{Dick}.  Due to the nearby $\pi\Delta$ 
threshold, both P$_{11}$ poles are not far from physical region 
(Fig.~\ref{fig1}).  There is a small shift between pole positions on the two 
sheets, due to a non-zero jump at the $\pi\Delta$-cut.  Our 
conclusion is that a simple BW parametrization cannot account for 
such a complicated structure. This point was also emphasized by 
H\"ohler~\cite{PDG}.  Recent studies by the J\"ulich\cite{Juelich} 
and EBAC\cite{Sato} groups have confirmed the two pole determination.
An earlier study by Cutkosky and Wang came to a similar conclusion~\cite{Cutk}.

Following the first indications from PWA studies, evidence~\cite{bnl} for
the Roper was found through the analysis of hydrogen bubble chamber events.
More recent evidence for a direct measurement of the N(1440) has been found
using electromagnetic interactions (at BES in $e^+e^-\to J/\psi\to p\pi^-\bar{n} + 
n\pi^+\bar{p}$\cite{BES} and at JLab in $ep\to e^\prime X$\cite{RSS}). 
Hadronic  processes (at SATURNE~II in $\alpha p\to\alpha^\prime X$\cite{SATURNE} 
and at Uppsala in $pp\to np\pi^+$\cite{WASA}) have also studied.  Some of the
peaks found have positions different from the BW interpretation of $\pi$N elastic 
scattering\cite{PDG,piN} with ``masses" closer to the real part of the pole 
position\cite{piN,piNo}.  These differences could reflect the complicated structure 
described above.

\begin{center}
\includegraphics[width=8cm, angle=0]{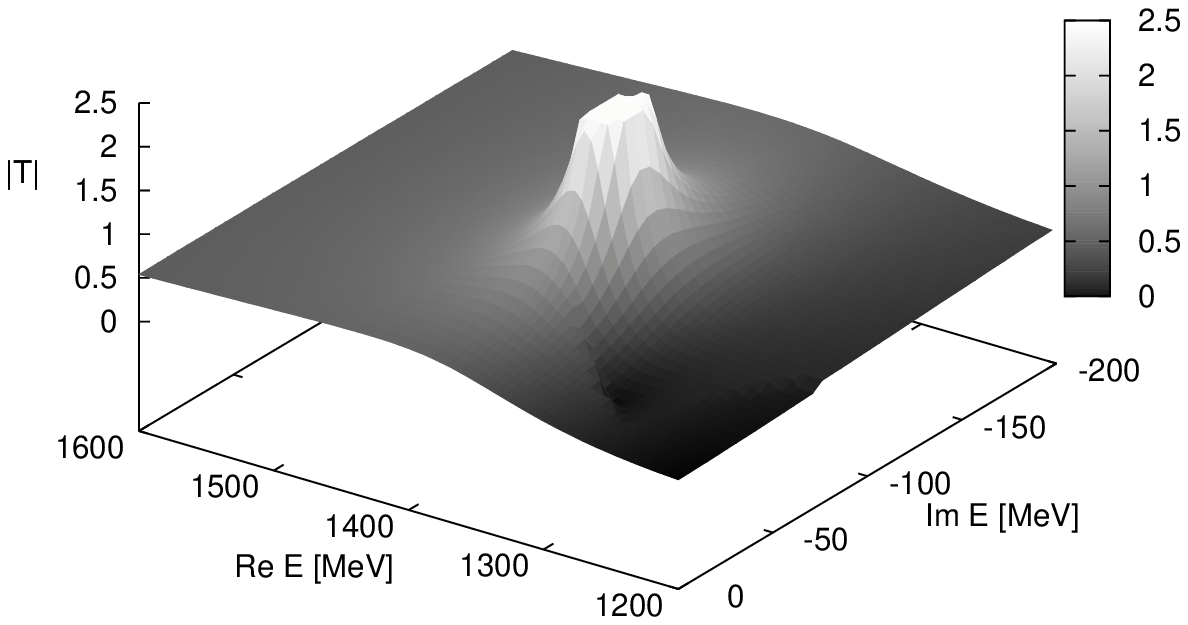}\hfill
\includegraphics[width=8cm, angle=0]{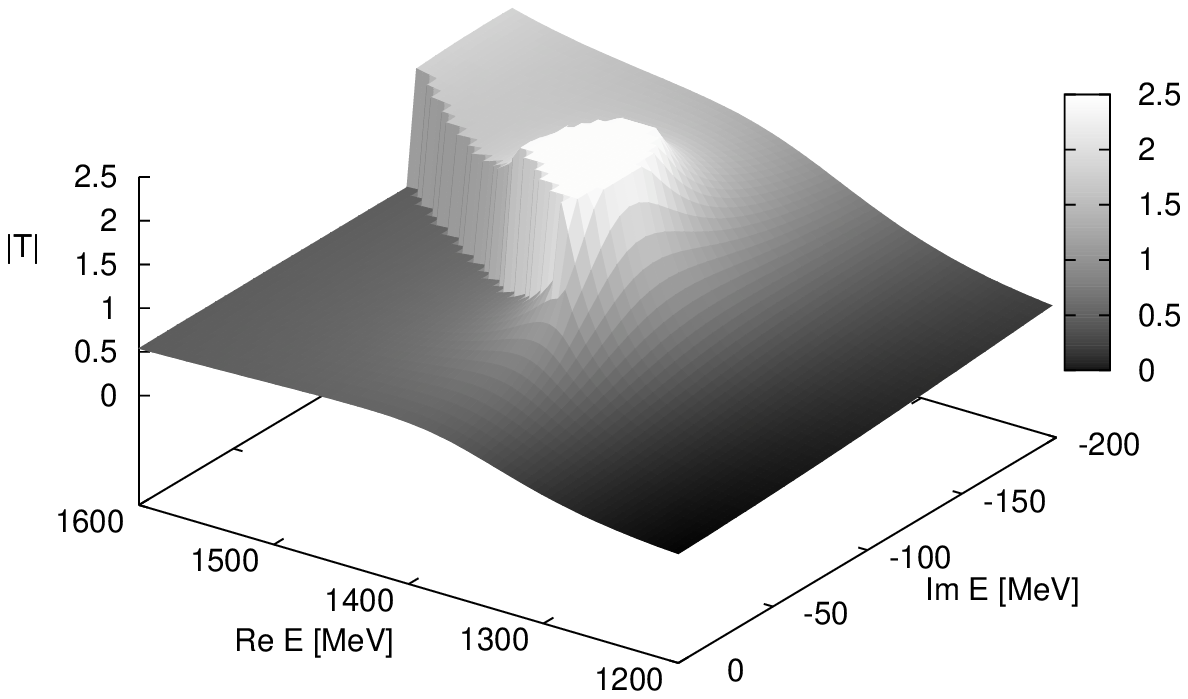}
\vspace{3mm}
\figcaption{\label{fig1} Two poles for $\pi$N P$_{11}$.
           Top: the $\pi \Delta$ cut can be seen in the
	   foreground and runs from larger to smaller values 
	   of the real part of the energy.
           Bottom: the $\pi \Delta$ cut is clearly visible
	   running from smaller to larger values of the
	   real part of the energy.}
\end{center}
Overall, most of analyses of N(1440) are based on its BW 
parametrization, which implicitly assumes that the resonance is related to an 
isolated pole.  However, given the complicated structure found in our PWA, 
the BW description may be  only an effective parametrization, which could 
be different in different processes.  Some inelastic data 
indirectly support this point, giving N(1440) BW masses and 
widths significantly different from the PDG BW values\cite{PDG}.  
This may also cast some doubt on recent $Q^2$ evaluation results\cite{MAID,CLAS}, 
since the $Q^2$-dependences for contributions of 
different singularities may be different.
This problem can be studied in future measurements with JLab CLAS12.

\subsection{P$_{11}$ beyond 1500~MeV}

Beyond the Roper resonance, the $P_{11}$ partial wave wraps around the center of the
Argand diagram (Fig.~\ref{fig2}) and the total elastic cross section
is half the total cross section (Fig.~\ref{fig3}).  As a result, small 
changes in the amplitude can produce large changes in the phase, 
though these changes have little influence on the fit to data.  For 
$\pi$N elastic scattering, we conclude that there is little 
sensitivity to resonances in P$_{11}$ above 1500~MeV except possible 
states with small $\Gamma_{el}$\cite{p11}.    

\begin{center}
\includegraphics[width=5cm, angle=90]{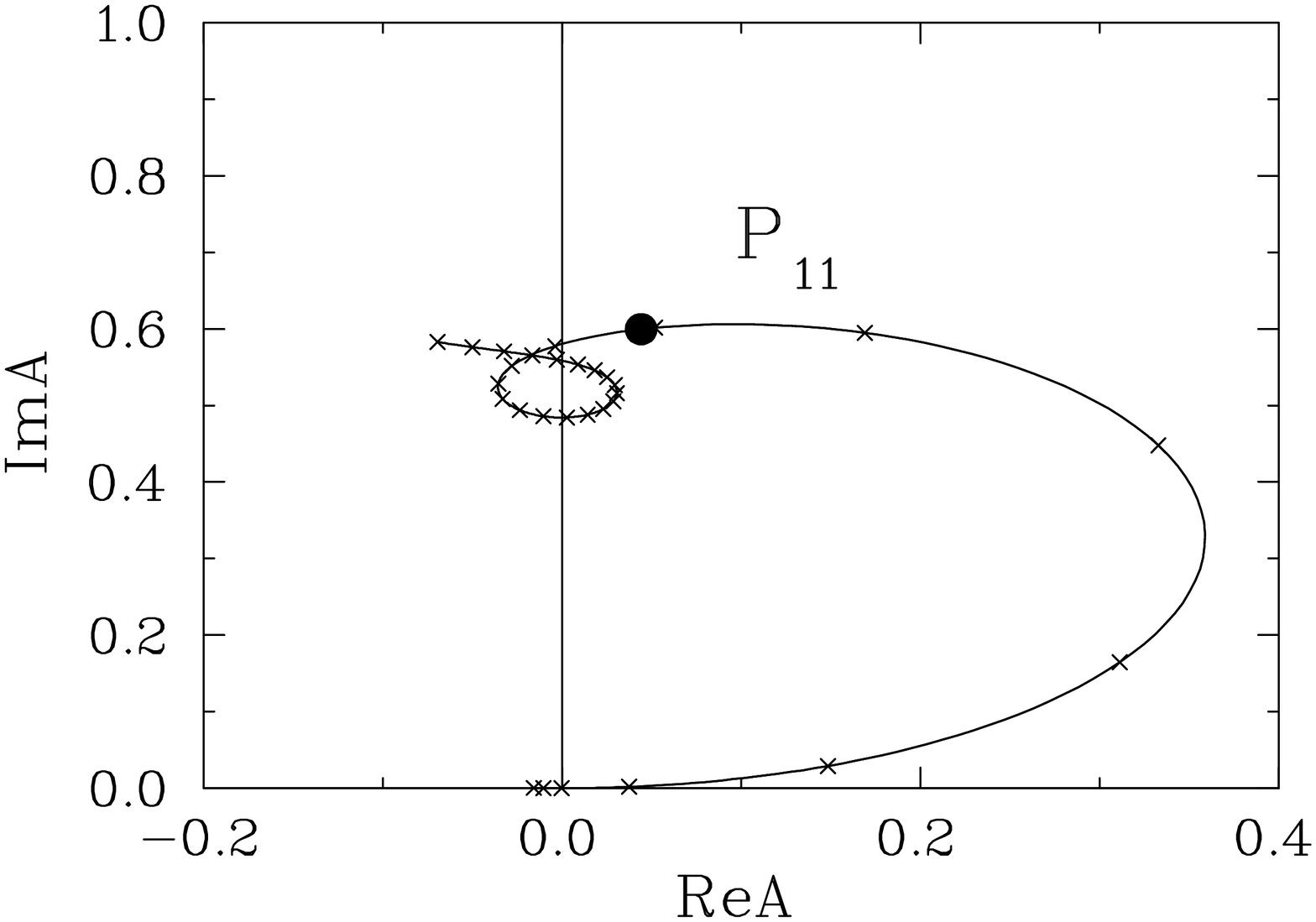}
\vspace{3mm}
\figcaption{\label{fig2} Argand plot for the $P_{11}$ partial-wave amplitude from 
         threshold (1080~MeV) to W = 2500~MeV.  Crosses indicate 
         50~MeV steps in W.  The solid circle corresponds to the SP06 BW 
         $W_R$.}
\end{center}
\begin{center}
\includegraphics[width=5cm, angle=90]{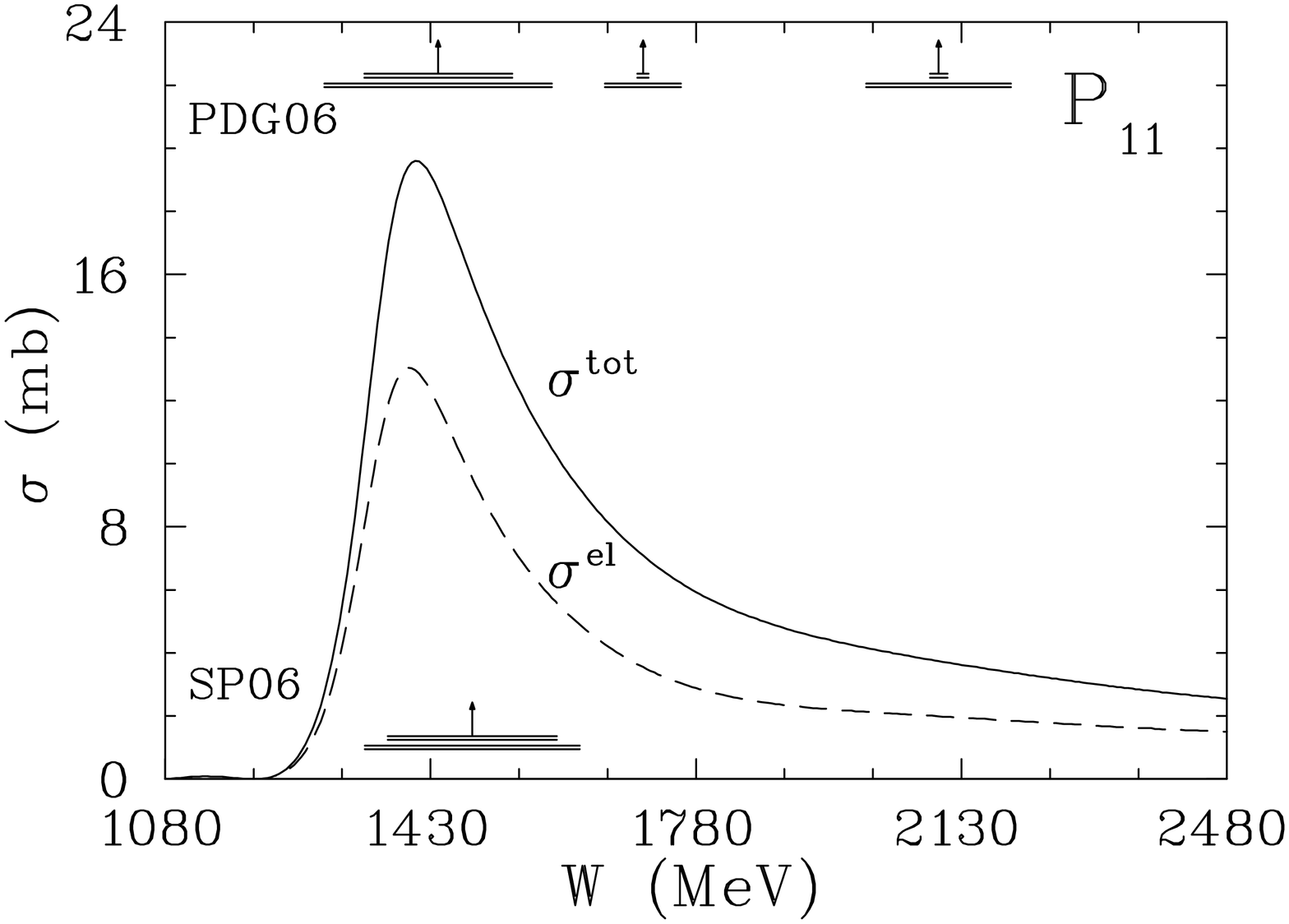}
\vspace{3mm}
\figcaption{\label{fig3} P$_{11}$ contribution to total and total 
          elastic cross sections for SP06.  Vertical arrows indicate 
          resonance $W_R$ values and horizontal bars show full $\Gamma$ 
          and partial $\Gamma_{\pi N}$ widths.  The lower BW resonance 
          symbols are associated with the SP06 values; upper symbols 
          give PDG values, which include higher mass states.}
\end{center}
One may speculate about the existence of a very narrow $P_{11}$ state which,
as mentioned above, would not be clearly evident in a standard PWA. Such a 
state was originally motivated by investigations aiming to explain how a very
narrow (less than 1~MeV) pentaquark state could exist. Here we can 
summarize our knowledge of one such ``narrow" candidate, N(1680)P$_{11}$:\\
(i)   Using a modified PWA\cite{p11}, designed to search for slots where a
      very narrow state would not destroy the existing fit to pion-nucleon 
      elastic scattering data, a candidate energy was found 
      at 1680~MeV with a $\Gamma_{\pi N} <$0.5~MeV.\\
(ii)  There are several independent suggestions for the
      N(1680)\cite{Graal,CB-Elsa,Lns}, \\
(iii) Its width is much less than any non-strange N$^\ast$\cite{p11,Graal,CB-Elsa,Lns}, \\
(iv)  The Chiral-soliton approach gives support for N(1680) 
      production in both $\gamma p$ and $\gamma n$\cite{Max},\\
(v)   The GRAAL $\gamma n\to\eta n$ cross section measurements allow one
      to determine the radiative width of N(1680) and transition 
      magnetic momentum\cite{Yakov} which is much smaller than for
      the $\Delta$ case.\\

\subsection{$\pi^-p\to\eta n$ Database Puzzle}

Most measurements of the $\pi^-p\to\eta n$ reaction cross section are 
rather old and sometimes conflicting (Fig.~\ref{fig4}).  There are few 
cross section (106 data) measurements above 800~MeV and no polarized 
measurements below 1040~MeV~\cite{said}.  A detailed analysis of the 
older data can be found in the review by Clajus and Nefkens~\cite{clnef}. 
Most NIMROD data do not satisfy a consistency requirement [systematics are not under
control, momentum uncertainties up to 50 -- 100~MeV/c, and so on]. 
For this reason, we are not able to use these data in our $\pi^-p$ elastic, 
$\pi^-p\to\pi^0n$, and $\pi^-p\to\eta n$ analyses of scattering data. 
In particular, the data above 800~MeV does not permit a model-independent 
analysis of $\pi^-p\to\eta n$.
\begin{center}
\includegraphics[width=5cm, angle=90]{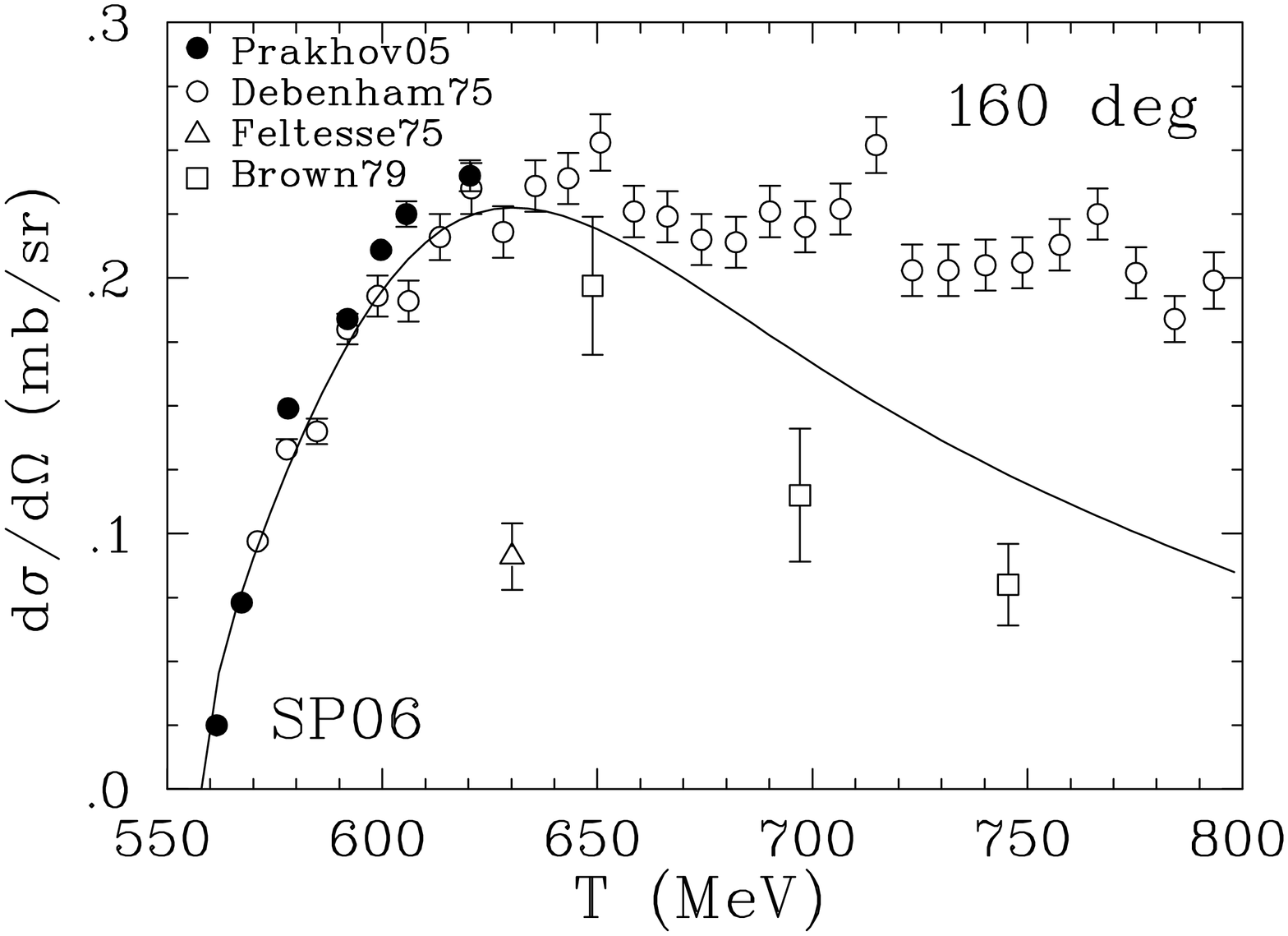}
\vspace{3mm}
\figcaption{\label{fig4} Fixed angle excitation functions for
           $\pi^-p\to\eta n$. We are not able to use data shown
           by open symbols in our analysis.}
\end{center}
The existing data types and energy limits severely restrict any attempt to determine 
resonance parameters above the first $S_{11}$ resonance.

\section{Pion Photo- and Electroproduction}

In fitting the electroproduction database, we extrapolate from the 
relatively well determined $Q^2$ = 0 point.  The photoproduction 
multipoles can be parametrized using a form containing the Born terms 
(no free parameters) and phenomenological pieces maintaining the 
correct threshold behavior and Watson's theorem below the two-pion 
production threshold.  The $\pi$N $T$ matrix connects each multipole to 
structure found in the elastic scattering analysis. The parametrization
above two-pion production is based on a unitary $K$-matrix approach, which
no strong constraints on the energy dependence apart from correct
threshold properties.  

Overall, the difference between MAID and GW/VPI amplitudes tends to 
be small but resonance content may be essentially different (Figs.~7 
and 8 from\cite{piPRp}).  One reason for differences is database dependent. 
MAID07\cite{MAID} did not use recent CLAS $\pi^0p$\cite{piPRn} and 
$\pi^+n$\cite{piPRp} with LEPS $\pi^0p$\cite{LEPS} backward measurements. 
Other differences
are tied to different assumptions regarding the inclusion of resonance
and background contributions. Some rather large differences are evident
in those wave connected to the  pion-nucleon $S_{11}$ and $D_{13}$ partial waves.

There are several issues in pion photoproduction above the 
$\Delta$(1232) which require resolution. We consider them in the
remainder of this section.

\subsection{Forward $\pi^0p$ Photoproduction}

For incident photon energies up to 1.3~GeV, the $\pi^0p$ data 
obtained the CLAS Collaboration\cite{piPRn} are for the most part 
in very good agreement with previous measurements. At higher 
energies, a disagreement between the CB-ELSA\cite{Bonn} measurements 
and the CLAS appears especially at forward angles (Fig.~8 
from Ref.\cite{piPRn}).  The overall systematic uncertainty for the 
CB-ELSA measurements is stated to be 5\% below 1300~MeV and 15\% 
above that energy.  This compares with the roughly 5\% systematic 
uncertainty obtained at JLab.

Moreover, the CLAS $\pi^0p$ measurements and SAID fit do not confirm 
the existence of weak states reported by the BoGa group in a fit to 
the CB-ELSA data\cite{BoGa}.  

Given the smooth behavior exhibited by the excitation functions in 
Figs.~9 and 10 from Ref.\cite{piPRn}, the CLAS cross sections provide 
no hint of ``missing" resonance structure between 2 and 3~GeV.  The 
SAID fits implicitly contain only those resonances found in the 
corresponding SAID analysis of elastic $\pi$N scattering data. No 
change in the form of the SAID photoproduction fit was found to be 
necessary. In contrast, the CB-ELSA fit required many additional 
resonance contributions, some of which are 1- and 2-star rated PDG 
states, as well as a new N(2070) resonance. One possible explanation 
is apparent in Fig.~10 from\cite{piPRn}, which shows the CLAS data 
to be somewhat smoother than the CB--ELSA excitation functions. 
Model-dependence in the separation of resonance and background 
contributions is also a critical factor.  This uncertainty may be 
reduced through measurements of further (polarized) data. 

Clearly, additional measurements at forward angles are needed to 
determine whether the rapid increase suggested by the most forward 
CB-ELSA data is correct, or whether the behavior suggested by the 
most recent fits properly describes the cross section at forward 
angles.  That is critical because the forward measurements are 
sensitivity to highest N$^\ast$s (most of these are inelastic).

\subsection{$\pi^-p$ Photoproduction}

Complementary measurements of $\pi^\pm$ photoproduction are required for 
an isospin decomposition of the multipoles.  There are no prior 
comprehensive tagged $\pi^-p$ measurements. Final-state-interactions (FSI) 
play a critical role in a state-of-the-art analysis of the $\gamma 
n\to\pi^-p$ data.  A preliminary study suggests FSI (Fermi motion 
included) varies 
between 15 and 40\% for the CLAS energy range (E$_\gamma$ = 1050 -- 
3500~MeV) and depends on the energy and scattering angle.  There 
are some previous measurements coming from hadronic facilities but few 
data are available to accomplish a reliable PWA and determine neutron 
couplings.  A JLab analysis addressed to these data is coming from the 
$\gamma d\to\pi^-pp$ experiment (g10 run period) ( in progress).  
The difference between previous and CLAS measurements may 
result in significant changes for the neutron couplings.

\subsection{Pion Electroproduction}

Ongoing fits incorporate all available electroproduction data, with 
modifications to our fitting procedure implemented as necessary 
(Table~\ref{tab2}).  We note that the CLAS Collaboration
produced 85\% of the world pion electroproduction data, much of which
was focused on the mapping of the properties of the 
$\Delta(1232)$ resonance.  Useful 
comparisons will require those involved in this effort to make 
available all amplitudes obtained in any new determination of $R_{EM}$ 
and $R_{SM}$ for the $\Delta$(1232) which may be compared with 
LQCD calculations\cite{lqcd}.  
\begin{center}
\tabcaption{ \label{tab2}  GW N$^\ast$ Program}
\footnotesize
\begin{tabular*}{80mm}{c@{\extracolsep{\fill}}cc}
\toprule Reaction       &  Data   & $\chi^2$ \\
\hline
$\gamma^\ast p\to\pi^0p$&  55,766 &  81,284 \\
$\gamma^\ast p\to\pi^+n$&  51,312 &  80,004 \\
Redundant               &  14,772 &  17,375 \\
Total                   & 124,453 & 178,663 \\
\hline    
$\gamma p\to\pi N$      &  24,888 &  50,684 \\
All Photo              & 159,341 & 229,317 \\
\hline    
$\pi N\to\pi N$         &  31,876 &  57,255 \\
\hline    
All $\pi N$             & 191,217 & 286,572 \\
\bottomrule
\end{tabular*}
\end{center}

Of all resonances, one might assume that the $\Delta$(1232) properties 
are know to great precision.
Unfortunately, this is not really true (Fig.~\ref{fig5}).  The PDG 
average values look stable while our determination depends on the
database.  The first jump (1990 -- 1993) is associated with the 
$\pi^0p$ LEGS activity, the 2nd jump (1993 -- 1996) is the 
product of the MAMI-B for $\pi^0p$ and Bonn $\pi^+n$ activity, 
the 3rd jump (1996 - 1997) depends again from MAMI-B for $\pi^0p$ 
and Bonn $\pi^+n$ activity, then 4th jump (1997 -- 2003) is the 
result of MAMI-B for $\pi^0p$ and Bonn with MAMI-B $\pi^+n$ 
activity, and finally,the 5th jump (2003 -- 2007) depends from MAMI-B 
for both $\pi^0p$ and $\pi^+n$. 
\begin{center}
\includegraphics[width=5cm, angle=90]{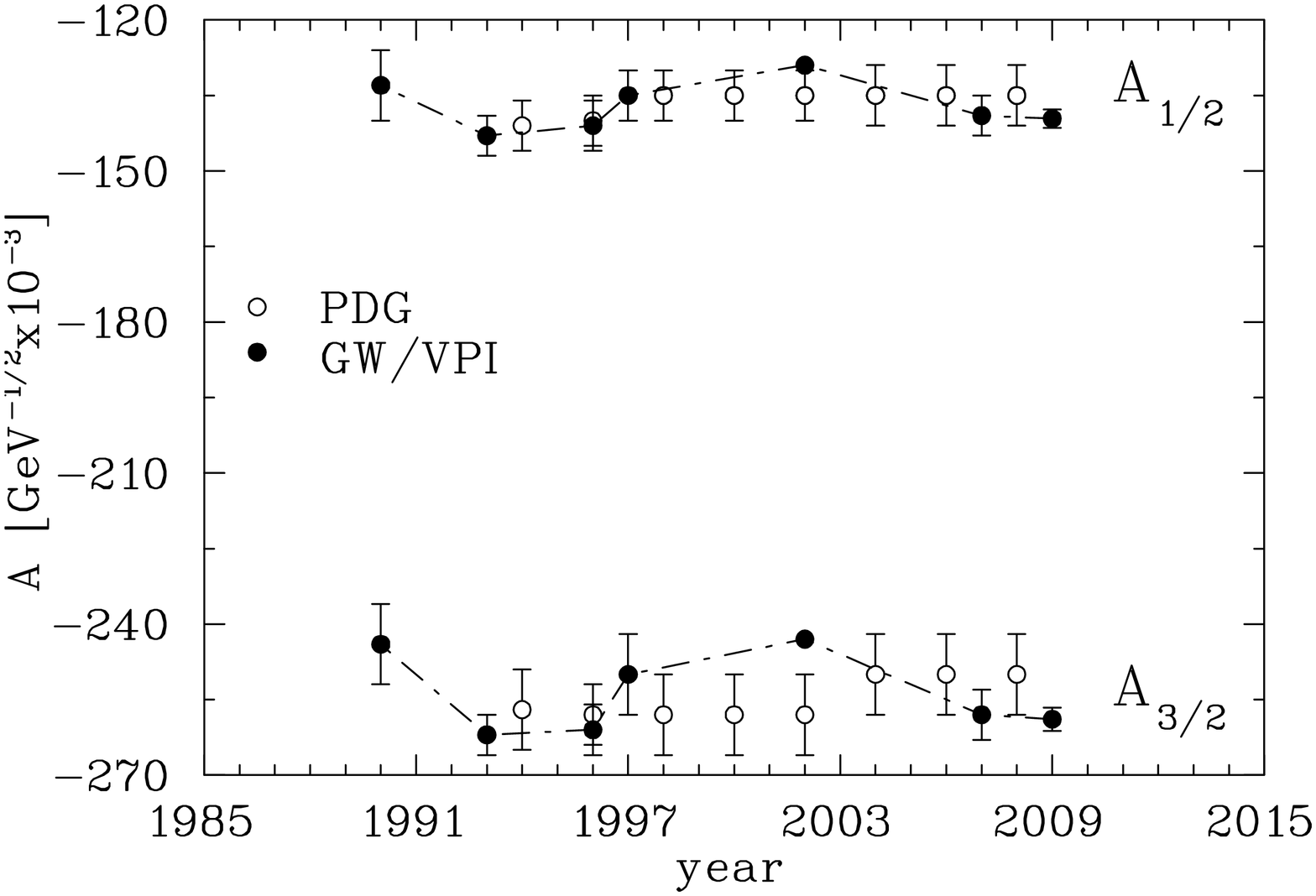}
\vspace{3mm}
\figcaption{\label{fig5} Time variation of the A$_{1/2}$ and 
           A$_{3/2}$ proton coulpings for the $\Delta$(1232).}
\end{center}

A major pion electroproduction database problem is that most data 
are from unpolarized measurements.  There are no $\pi^0n$ data 
and very few $\pi^-p$ data (no polarized measurements).  This 
does not allow a rigorous neutron coupling evaluation vs. Q$^2$.  
The Q$^2$ distribution of available data is shown in
Fig.~\ref{fig6}
\begin{center}
\includegraphics[width=4cm, angle=90]{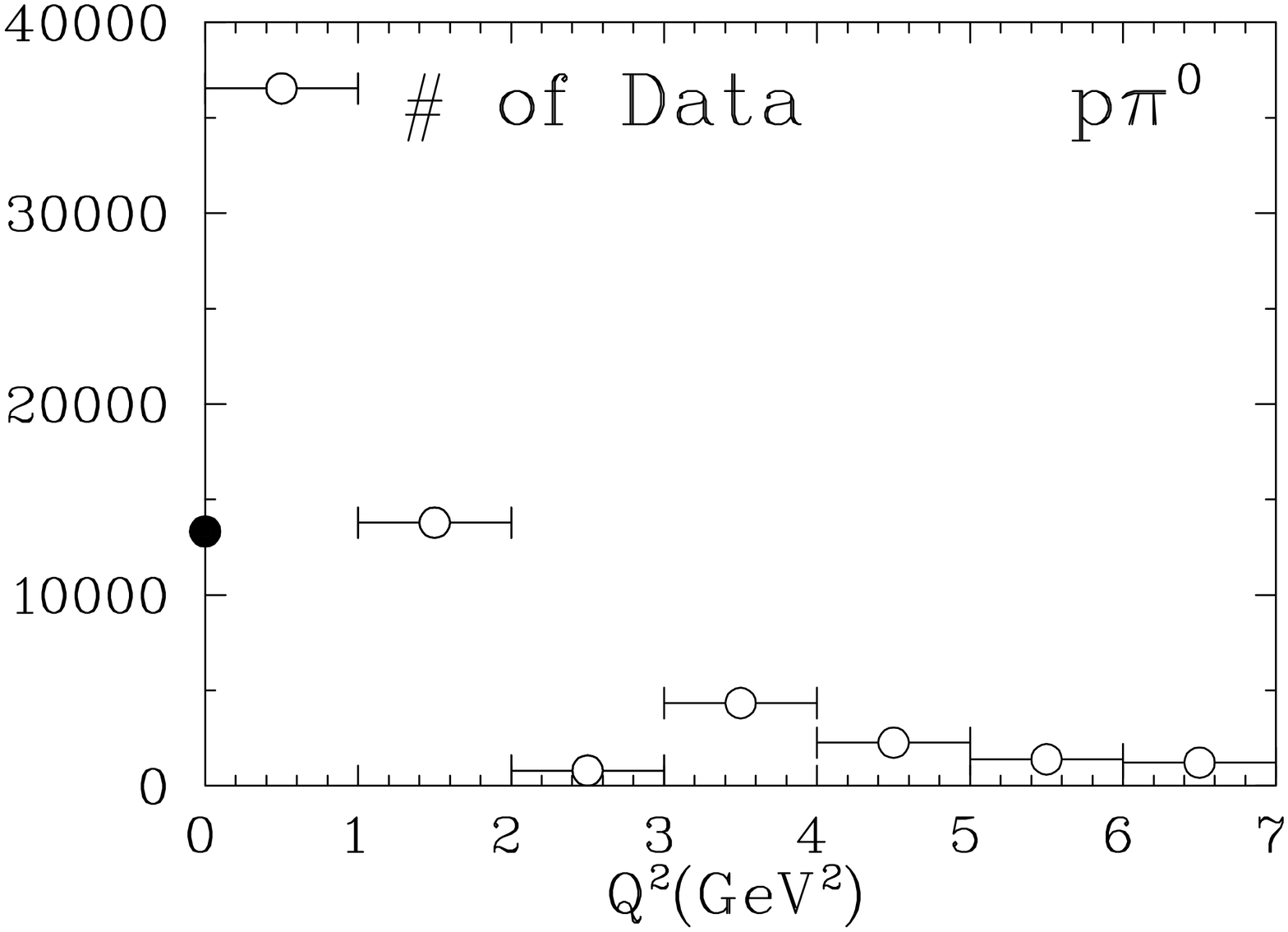}\hfill
\includegraphics[width=4cm, angle=90]{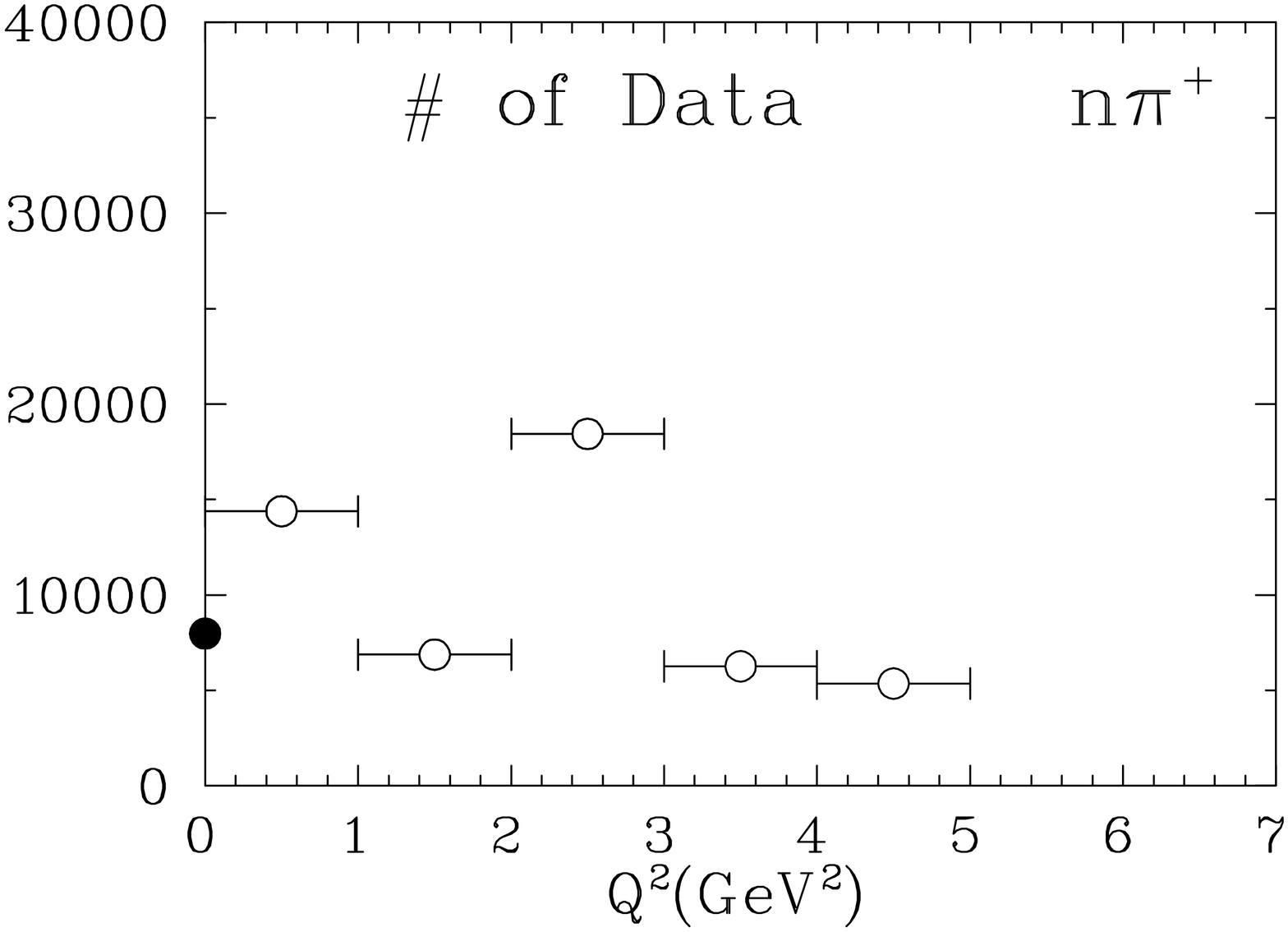}\hfill
\includegraphics[width=4cm, angle=90]{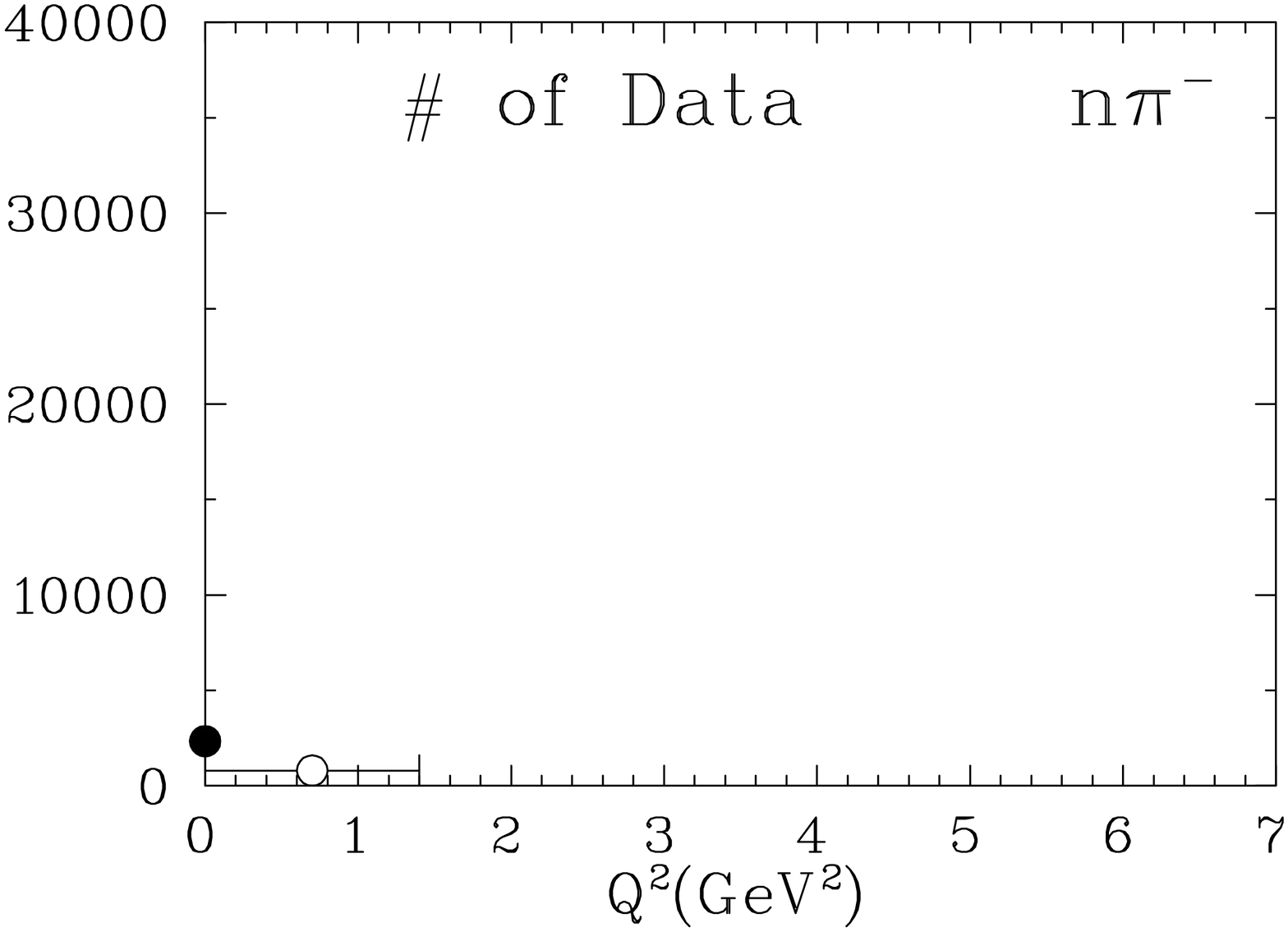}
\figcaption{\label{fig6} Q$^2$ distribution of pion
           electroproduction data which are now available.}
\end{center}

\section{Summary and Prospects}

Let us, in the interest of clarity, summarize where the analysis of
the single meson productions reactions stands.\\
i)   $\pi N$ analysis is crucial for the N$^\ast$ program, \\
ii)  Extended $\pi$N elastic and pion production analyses are done 
     up to W = 2500~MeV,\\
iii) $\pi N\to\eta N$ and eta photoproduction analyses are done 
     up to W = 1640~MeV.\\

Looking forward, our efforts will be focused on the following important
issues.\\
i)   Production measurements on the ``neutron" target are necessary
     to determine neutron couplings at Q$^2$ = 0,\\
ii)  Future improvement will be possible with future measurements of 
     spin observables at JLab, MAMI-C, LEPS, LNS, and CB-ELSA,\\
iii) Complete experiments make possible a direct reconstruction of 
     helicity amplitudes for pion and eta photoproduction.\\

Finally, issues which will receive further attention are as follows.\\
i)   Q$^2$ evaluation of resonance couplings up to very large Q$^2$,\\
ii)  The critical question is can we reach an asymptotic regime as 
     pQCD predicted?\\
iii) Neutron electroproduction measurements are necessary to determine 
     neutron couplings at Q$^2 >$ 0.\\



\end{multicols}

\end{document}